\documentclass[aps,twocolumn,prb,tightenlines,floatfix,showpacs,superscriptaddress]{revtex4-1}
\usepackage{amssymb}
\usepackage{amsmath}
\usepackage{mathrsfs}
\usepackage{graphicx}

\newcommand{\kt}[1]{\ensuremath{|#1 \rangle}}

\newcommand{\bs}[1]{\boldsymbol{#1}}

\newcommand{\pmt}[1]{\begin{pmatrix} #1 \end{pmatrix}}

\newcommand{\pth}[1]{\left( #1 \right)}
\newcommand{\brc}[1]{\left\{ #1 \right\}}
\newcommand{\bckt}[1]{\left[ #1 \right]}
\newcommand{\abs}[1]{\left| #1 \right|}
\newcommand{\avg}[1]{\left\langle #1 \right\rangle}

\newcommand{\upa}[0]{\uparrow}
\newcommand{\dna}[0]{\downarrow}
\newcommand{\ria}[0]{\rightarrow}

\newcommand{\inv}[0]{{}^{-1}}

\newcommand{\trm}[1]{{\textrm{#1}}}
\newcommand{\mrm}[1]{{\mathrm{#1}}}
\newcommand{\mcl}[1]{{\mathcal{#1}}}
\newcommand{\mbb}[1]{{\mathbb{#1}}}

\newcommand{\meq}[1]{\begin{equation} #1 \end{equation}}
\newcommand{\mea}[1]{\begin{align} #1 \end{align}}

\begin{document}

\title{General pairing theory for condensed and non-condensed pairs of a 
superconductor in a high magnetic field}

\author{Peter Scherpelz}
\author{Dan Wulin}
\author{B{\v r}etislav {\v S}op{\'\i}k}
\author{K. Levin}
\affiliation{James Franck Institute and Department of Physics,
University of Chicago, Chicago, Illinois 60637, USA}
\author{A. K. Rajagopal}
\affiliation{Inspire Institute Inc., Alexandria, Virginia 22303, USA}
\affiliation{Harish-Chandra Research Institute, Chhatnag Road, Jhunsi,
Allahabad, 211019, India}

\date{\today}

\begin{abstract}
We extend Gor'kov theory to address
superconducting pairing at high magnetic fields and general
temperatures with arbitrary attractive
interaction strength.
This analysis begins with
a new interpretation of the high-field Gor'kov gap equation
which we associate with an instability in  
a generalized particle-particle ladder series. 
Importantly, this interpretation of the non-linear gap equation
enables a treatment of pairing which
is distinct from condensation.
We also show how to consolidate two distinct fermionic pairing schemes
in real and momentum space,
both corresponding to an Abrikosov lattice.
Numerical results for the fermionic local density of states
demonstrate that gapless structure in a field is robust and presumably relevant
to quantum oscillation experiments. We find that despite their differences,
both pairing
schemes contain very similar physics.
Our formalism is designed to explore
a variety of
magnetic field
effects in the so-called pseudogap phase and throughout the
BCS-BEC crossover.
\end{abstract}

\maketitle

\section{Introduction }

High magnetic field superconductivity is a difficult theoretical problem with
important implications for experiment.
Observations of
quantum oscillations in the superconducting phase of conventional
superconductors were initially
unexpected,\cite{maniv_2001} and further unusual oscillations have been observed
in underdoped cuprates.\cite{Taillefer3} 
At the same time, theoretical treatments 
reached surprising conclusions, with some
investigations finding re-entrant superconductivity at extremely high magnetic
fields.\cite{rasolt_1992} Adding to the complexity is the observation
that the introduction of Landau levels into a pairing scheme 
appears to cause
a three-dimensional
superconductor to behave like a one-dimensional system.
This greatly enhances fluctuation effects \cite{lee_1972}
and may even de-stabilize
superfluid condensation.

The goal of the present paper is to set up a foundation for
addressing these issues by extending the standard Gor'kov
(BCS-based) approach to apply to
general temperatures $T$, away from the instability
regime, and to stronger-than-BCS
attractive interactions. 
Although the immediate focus of this paper is on the ordered phase,
we use the Gor'kov theory extension 
to arrive at a compatible description of
the normal phase as well. Importantly, this normal phase may
possess a rich structure associated with precursor pairing in
the presence of magnetic fields.

In a related paper \cite{scherpelz_2012}
we focus on this disordered phase and explore the notion that some degree of
pairing at the onset of condensation may
be necessary
to avoid
a strict dimensional reduction that prohibits condensation into a
superconducting state altogether.  Furthermore, these
excited pair states may be present in systems
such as high temperature
superconductors or in fermionic gases in the
BCS-BEC crossover regime, where the non-condensed pairs 
are associated with a pseudogap
state.
It should be stressed
that
the non-condensed pairs we consider are distinct from conventional fluctuations.  These pairs
arise from strong attractive interactions, not from low dimensionality (and/or
disorder effects) which give rise to conventional fluctuations.
Our BCS-BEC based approach is similarly distinct from the so-called
``phase fluctuation'' scenario which is based on soft phase fluctuations
presumably arising from low plasma frequencies. Indeed, since we are
contemplating both charged and uncharged superfluids, the issue of
soft plasma frequencies is not particularly generic.

Gor'kov theory addresses the fermionic degrees of freedom. 
Two
proposals\cite{ryan_1993,dukan_1991} have been put forth to describe
the nature of those fermionic
pairs which form the condensate in the presence of high magnetic fields. 
These are associated with orbit center-based and magnetic lattice-based pairing
schemes. Here we show how the physical implications of each are similar
and that both lead to gapless fermionic states which are thought to
be the basis for observed quantum oscillations.\cite{maniv_2001}
A central contribution of the present paper is to demonstrate that
the (analytically tractable) theoretical structure of these different pairing
approaches can be consolidated into a more general
formulation which addresses the \textit{non-linear} structure
of the Gor'kov theory. 
This is in contrast to a substantial fraction of the literature on
high magnetic field superconductivity which deals with the linear regime
where the gap is small.
We thus arrive at an interpretation of the Gor'kov gap equation
which allows us to extract a set of particle-particle ladder diagrams
which properly characterize the pairing fluctuations or non-condensed pairs
in the presence of high magnetic fields.

We begin in Section \ref{secLLRef} by deriving the Landau level representation
of the Gor'kov equations.  In Section \ref{secIntra} we discuss a
``diagonal"
approximation made to these equations which is thought 
\cite{tesanovic_1998}
to
be suitable for high magnetic
fields.  Following this, in section \ref{secLadder} we focus on the gap equation and discuss its
relationship to a divergent particle-particle ladder series.  In section
\ref{secNoncon} we show that this divergent series captures the contribution of
non-condensed pairs to the theory.

The second half of the paper is less general and more
concrete. Here we focus on implementing and comparing 
different state-space representations of this pairing theory. In Section
\ref{sec:pairing} we discuss the two different existing implementations.  In
Section \ref{secNNPair} we show how to use a ``tight-binding'' approximation to
orbit-center pairing to make the theory analytically tractable, facilitating
comparison of the two implementations.  
Finally, in Section \ref{secResults} we show how
gaplessness is robust among these pairing theories by comparing local
density of states calculations.  Our conclusions
are presented in Section \ref{secConclusion}.

\section{Derivation of Gor'kov's Equations in a Landau Level
Basis\label{secLLRef}}
We begin with the Gor'kov equations formulated in real space in terms of the gap
$\Delta(\bs r)$
and Green's functions in integral form:
\mea{G(&\bs r,\bs r';
i\omega) = G_0(\bs r,\bs r'; i\omega) - \int d\bs r'' d \bs r''' G_0(\bs r,\bs
r'';i\omega) \notag \\
&{}\times \Delta(\bs r'')G_0(\bs r''',\bs r''; -i\omega) \Delta^\dag(\bs
r''')G(\bs r''',\bs r'; i\omega)\label{GorkovGR}} 
\meq{\Delta^\dag(\bs r) = -
\frac{g}{\beta}\sum_{i\omega}\int d\bs r'G(\bs r',\bs r;i\omega)G_0(\bs r',\bs
r; -i\omega)\Delta^\dag(\bs r')\label{GorkovGapR}} 
where $g$ is the interaction
strength, and $i\omega$ ($i\Omega$) denote fermionic (bosonic) Matsubara
frequencies with the traditional subscripts omitted for clarity.  
These equations allow the identification of the self-energy $\Sigma$, 
\meq{\Sigma(\bs
r,\bs r';i\omega) = -\Delta(\bs r)\Delta^\dag(\bs r ')G_0(\bs r',\bs
r;-i\omega),\label{GorkovSigmaR}} 
and we now rewrite Eqs.~(\ref{GorkovGR}-\ref{GorkovSigmaR}) in terms of the
Landau level representation that diagonalizes the 
non-interacting Hamiltonian
$\mcl{H}_0$.
\cite{vavilov_1997}
The bare Green's function $G_0$ is then given by
\mea{G_0(\bs r,\bs r';i\omega) &= \sum_n
G^0_n(i\omega)\psi_n(\bs r)\psi_n^\dag(\bs r') \notag \\ 
&= \sum_n \frac{\psi_n(\bs
r)\psi_n^\dag(\bs r')}{i\omega-\xi_n}} 
where $n = (N,p,k_z)$ is the Landau level state, with $N$ the Landau
level, $p$ the degenerate Landau level index, $k_z$ the momentum in the
$z$-direction (parallel to a magnetic field $\bs B$), and $\xi_n$ the energy
of a particle in state $n$.  $G$, however, is not in general diagonal in the
Landau level representation, and is given by
\meq{G(\bs r,\bs r';i\omega) =
\sum_{mm'}G_{mm'}(i\omega)\psi_m(\bs r)\psi^\dag_{m'}(\bs r').\label{eq:Gr}} 
In this representation, Eq.~\eqref{GorkovGR} is multiplied by
$\psi_{m}^\dag(\bs r)\psi_{m'}(\bs r')$ and integrated over both $\bs r$ and
$\bs r'$ to give 
\mea{G_{mm'}&(i\omega) = G^0_m(i\omega)\delta_{mm'} - \sum_{l
n}\int d\bs r'' d\bs r''' G^0_{m}(i\omega) \notag \\
&{}\times\psi^\dag_{m}(\bs r'')\Delta(\bs
r'')G^0_l(-i\omega)\psi_l(\bs r''')\psi_l^\dag(\bs r'')\Delta^\dag(\bs
r''') \notag \\
&{}\times G_{nm'}(i\omega)\psi_n(\bs r'''). \notag}

We then define a ``state-space gap''
\meq{\Delta_{mn} \equiv \int d\bs r \Delta(\bs r)\psi_m^\dag(\bs
r)\psi_n^\dag(\bs r)\label{DeltaDef}} 
and obtain
\mea{G_{mm'}(i\omega) =&\ 
G^0_m(i\omega)\delta_{m m'} - \sum_{l n }
G^0_{m}(i\omega)\Delta_{m l}\notag \\
&{}\times G^0_l(-i\omega)\Delta^\dag_{l n}G_{nm'}(i\omega).\label{LLGFunc}}

We multiply the gap equation, Eq.~\eqref{GorkovGapR}, by $\Delta(\bs r)$,
and express the right hand side in the Landau level representation to find
\mea{\int &d\bs r \abs{\Delta(\bs r)}^2 = -\frac{g}{\beta}
\sum_{mm'n}\sum_{i\omega}
\int
d\bs r'd\bs r G_{mm'}(i\omega) \notag \\
&{}\times G^0_n(-i\omega)\Delta^\dag(\bs r')\Delta(\bs
r)\psi_m(\bs r')\psi^\dag_{m'}(\bs r)\psi_n(\bs r')\psi^\dag_n(\bs r). \notag}
Using the above identity for $\Delta_{mn}$, the gap equation becomes
\meq{1 = -\frac{g}{\beta}\sum_{m m'
n}\sum_{i\omega}\frac{\Delta_{m' n}\Delta^\dag_{mn}}{\int d\bs r \abs{\Delta(\bs
r)}^2}G_{mm'}(i\omega)G^0_n(-i\omega).\label{LLDeltaEqn}}

The self-energy is similarly expressed in the Landau level representation,
\meq{\Sigma(\bs r, \bs r';i\omega) = \sum_{mm'}\Sigma_{mm'}(i\omega)\psi_m(\bs
r)\psi^\dag_{m'}(\bs r').}  
Using Eq.~\eqref{GorkovSigmaR} for $\Sigma$,
we find
\meq{\Sigma_{mm'}(i\omega) =
-\sum_nG^0_n(-i\omega)\Delta_{mn}\Delta^\dag_{nm'}.\label{LLSE}}

In summary, the Gor'kov equations in the Landau level
representation for a constant magnetic field are 
\mea{G_{mm'}(i\omega) &= G^0_m(i\omega)\delta_{m
m'} - \sum_{l n } G^0_{m}(i\omega)\Delta_{m l}\notag \\
&\ {}\times G^0_l(-i\omega)\Delta^\dag_{l n}G_{nm'}(i\omega) 
\tag{\ref{LLGFunc}}}
\meq{1 = -\frac{g}{\beta}\sum_{m m' n}\sum_{i\omega}\frac{\Delta_{m'
n}\Delta^\dag_{mn}}{\int d\bs r \abs{\Delta(\bs
r)}^2}G_{mm'}(i\omega)G^0_n(-i\omega) \tag{\ref{LLDeltaEqn}}}
\meq{\Sigma_{mm'}(i\omega) =
-\sum_nG^0_n(-i\omega)\Delta_{mn}\Delta^\dag_{nm'}.\tag{\ref{LLSE}}}

\section{Intra-Eigenstate Pairing\label{secIntra}}
To make further analytical progress, we must simplify
Eqs.~(\ref{LLGFunc},\ref{LLDeltaEqn},\ref{LLSE}).  Thus, 
as in the literature 
\cite{ryan_1993,dukan_1994,vavilov_1997}
we assume 
that the pairing involves degenerate eigenstates.
This assumption is justified if we are in a regime where $\abs{\Delta}$ is much
less than the splitting between the Landau levels
$\hbar\omega_c$, where $\omega_c = eH/(\hbar c m)$ is the cyclotron frequency.
This will be a good approximation at very high fields where the splitting
between Landau levels is large, the regime we explore in this paper.  
This approach has been 
analyzed carefully in
Ref.~\onlinecite{tesanovic_1998} which showed that in this regime inter-Landau
level effects that we neglect are perturbative.

We also simplify by defining
\meq{\Delta^0_{mn}(\zeta) \equiv \frac{\Delta_{mn}(\zeta)}{\sqrt{\int d\bs r
\abs{\Delta(\bs r,\zeta)}^2}} \equiv
\frac{\Delta_{mn}(\zeta)}{\Delta}.\label{Deltamn}} 
Note that we have introduced an important parameter $\zeta$ which
labels
different functional forms of $\Delta(\bs r)$. Throughout this
paper we will associate the
particular value $\zeta_0$ with the condensate configuration
$\Delta^\mrm{sc}(\bs r,\zeta_0)$, which may be
distinct from configurations occupied by non-condensed pairs, discussed below. 
It is convenient to define
\meq{\phi^2_{mm'}(\zeta) \equiv \sum_{n}\Delta^0_{mn}(\zeta)
\Delta^{0\dag}_{nm'}(\zeta).}

In order that all the potential pairing partners $n$ of a state $m$ are
energy degenerate,
the index
$n$ is in the same Landau level as
$m$, and pairing occurs between states with $z$-momenta $k_z$ and $-k_z$.  This
allows us to write the fermionic self-energy Eq.~\eqref{LLSE} as 
\meq{\Sigma_{mm'}(i\omega) =
-G^0_N(k_z;-i\omega) \abs{\Delta}^2 \phi^2_{mm'}(\zeta_0),} 
where we write $G^0$ in terms
of only the Landau level $N$ and the $z$-momentum $k_z$ of $m$, with $N_m =
N_{m'}$ and $k_{z_{m}} = k_{z_{m'}}$.  This simplification of the gap
equation, Eq.~\eqref{LLDeltaEqn}, leads to
\meq{1 = -\frac{g}{\beta}\sum_{m m'}\sum_{i\omega}
\phi^2_{mm'}(\zeta_0)
G_{mm'}(i\omega)G^0_N(k_z;-i\omega). \label{eq:17}}

\section{The Gap Equation as Divergent Particle-Particle Ladder
Series\label{secLadder}}
\begin{figure*}[htbp]
\begin{center}
\includegraphics[trim=1.2in 8.8in 0in 1.2in,clip=true]{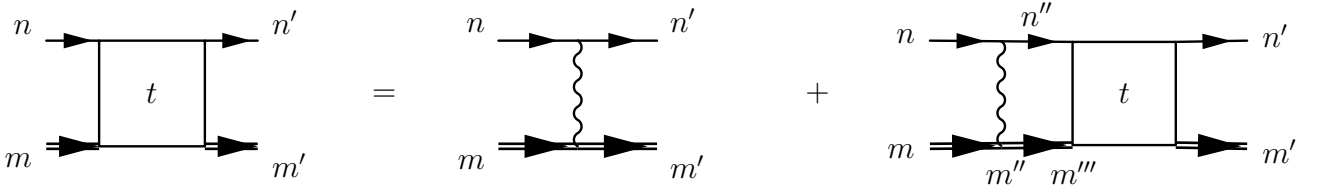}
\caption{The particle-particle ladder summation corresponding to a $t$-matrix
$t^\mrm{pg}$,
the divergence of which gives the general gap equation for the
superconducting state.}
\label{LadderSum}
\end{center}
\end{figure*}
The above formulation enables a consolidation of
standard approaches 
in the literature.\cite{ryan_1993,dukan_1991,dukan_1994,vavilov_1997}
In this context we re-interpret the gap equation of Eq.~(\ref{eq:17})
as a generalized
Thouless condition, which applies to all temperatures below
the transition. 
In the process we show that
this gap equation
serves to identify
a particular particle-particle ladder series. The
divergence of this series is a prerequisite for, and an indicator of, the
superconducting state.
Importantly this ladder series also
leads us to a characterization
of the associated non-condensed pairs, which may form above the critical
temperature for stronger-than-BCS
interactions.

It is essential first to characterize the
degrees of freedom available to these non-condensed pairs.
In the $z$-direction parallel to the magnetic field, the system behaves as in
zero-field, where condensed pairs are constructed of fermions with momenta $k_z$
and $-k_z$.  Thus, excited pairs must have nonzero total momentum, and we
can describe the general pairing of non-condensed pairs as being between momenta
$k_z$ and $-k_z+q_z$.\cite{chen_2005}

In the plane perpendicular to the magnetic field, the condensed electrons are
those which pair to form the real-space superconducting gap.
Here and throughout we distinguish the order parameter
$\Delta^\mrm{sc}
(\bs r,\zeta_0)$ from the excitation gap.
In a mean-field scheme, such as ours, 
where individual vortex fluctuations are not
included, we assume that
$\zeta_0$ corresponds to an Abrikosov lattice with functional form 
$\Delta^0(\bs r,\zeta_0) = \Delta^\mrm{sc}(\bs r,\zeta_0)/\Delta$.
We take the \textit{non-condensed} pairs to be those that form other
real-space gaps $\Delta^0(\bs r, \zeta)$ for $\zeta \neq \zeta_0$.  Finally, we
also allow the Matsubara frequencies to appear with total frequency
$i\Omega$.

We next introduce the pair susceptibility,
$\chi(\zeta,q_z;i\Omega) \equiv $ 
\meq{\frac{1}{\beta}\sum_{mm'}\sum_{i\omega}
\phi^2_{mm'}(\zeta)G_{mm'}(i\omega)G^0_N(k_z-q_z;i\Omega-i\omega)
.\label{chiEq}} 
With this important
definition, and the introduction of non-condensed pairs, we return to 
 the gap equation,
Eq.~(\ref{eq:17}), which can be rewritten as 
\meq{1 + g\chi \Big(\zeta = \zeta_0,q_z =0;i\Omega =0 \Big) =0  \label{eq:19}.}
One can interpret this equation as reflecting a 
divergence of a 
particle-particle ladder summation, shown in Fig.~\ref{LadderSum}.
We argue below that the gap equation is to be associated with a
$t$-matrix formed from the ladder diagrams in Fig.~\ref{LadderSum}, given here
in an abbreviated form by
\meq{t^\mrm{pg}(\zeta,q_z;i\Omega) 
= \frac{g}{1+g\chi(\zeta, q_z,i\Omega)}\label{eq:20}} (see Section
\ref{secNoncon} for details). 
For the condensate configuration, $\zeta = \zeta_0$, $q_z= \Omega = 0$,
the $t$-matrix thus diverges for all temperatures below the instability, as in
a Bose Einstein condensation condition, where the pairs have
vanishing chemical potential.
This ladder diagram set 
is to be distinguished from
a series which was previously identified to correspond to
the specific instability point
Ref.~\onlinecite{macdonald_1992}. Here, the condition is stronger, as
the incorporation
of one dressed $G$ and one bare
$G^0$ makes this summation valid throughout the superconducting regime.
Furthermore, as we explain in Section \ref{secNoncon}, it can be
extended to the entire strong attraction (BCS-BEC) regime where there is
a finite excitation gap at the instability, and in which pairing
and condensation must be distinguished.
\cite{scherpelz_2012,chen_2005}

\section{Characterizing the Non-condensed Pairs in the Gor'kov
Equations\label{secNoncon}}

BCS theory represents a very special case of superfluidity in which
pairing and condensation take place at the same temperature.
We have just argued that the
Gor'kov gap equation of Eq.~(\ref{eq:19}) 
and the closely related $t$-marix of Eq.~(\ref{eq:20}) effectively
constrain the nature of non-condensed pairs which below the 
transition condense into a
state with gap structure
$\Delta^0(\bs r,\zeta_0)$. 

For the moment, we consider Eq.~(\ref{eq:20})
as an appropriate characterization of the $t$-matrix (or effective
propagator) associated with the non-condensed pairs. We next
characterize their feedback into the Gor'kov equations, by
including $t$
in the self energy.
In a strict Gor'kov theory
$\Sigma^\mrm{sc}_{mm'}(i\omega) =$
\meq{\sum_{\zeta,q_z,i\Omega}\phi^2_{mm'}(\zeta)
t^{\mrm{sc}}_{mm'}(\zeta,q_z;i\Omega)
G^0_N(q_z-k_z;i\Omega-i\omega),}
where
\meq{t^\mrm{sc}(\zeta,q_z;i\Omega) \equiv
-\delta(\zeta-\zeta_0)\delta(q_z)\delta(i\Omega)\abs{\Delta^\mrm{sc}}^2,} with
$\Delta^\mrm{sc}$ corresponding to the gap $\Delta$ in Eq.~\eqref{Deltamn}.

We now include in the self-energy the non-condensed pair propagator $t^{pg}$
given by Eq.~\eqref{eq:20}, and in this way go beyond strict Gor'kov theory.  This
contribution is $\Sigma^\mrm{pg}_{mm'}(i\omega) = $
\meq{\sum_{\zeta,q_z,i\Omega}\phi^2_{mm'}(\zeta)t^{\mrm{pg}}
(\zeta,q_z;i\Omega)
G^0_N(q_z-k_z;i\Omega-i\omega).\label{Sigmapg}} 
Furthermore, since
$1+g\chi(\zeta_0,0;0)$ diverges below the
critical temperature,
$t^\mrm{pg}$
will be strongly peaked for 
parameters $(\zeta,q_z;i\Omega) \approx (\zeta_0,0;0).$
Since $i\Omega$ and $q_z$ will be
small for the primary
contributions to $t^\mrm{pg}$, we approximate the right hand
side using $G^0_N(q_z-k_z;i\Omega-i\omega)
\approx G^0_N(-k_z;-i\omega)$.

Then the total self-energy is now
\mea{\Sigma_{mm'}&(i\omega) = \Sigma^\mrm{sc}_{mm'}(i\omega) +
\Sigma^\mrm{pg}_{mm'}(i\omega) \notag \\
&= G^0_N(-k_z;-i\omega)\sum_{\zeta,q_z,i\Omega}
\phi^2_{mm'}(\zeta) \notag \\
&\ {}\times\pth{t^\mrm{sc}(\zeta,q_z;i\Omega) +
t^\mrm{pg}(\zeta,q_z;i\Omega)},} where $\Sigma^\mrm{sc}_{mm'}$ is the original
self-energy from Gor'kov theory defined in Eq.~\eqref{LLSE}, and 
$\Sigma^\mrm{pg}_{mm'}$ is
defined in Eq.~\eqref{Sigmapg}.

We can further simplify by defining a total $\abs{\Delta}^2$,
\meq{\abs{\Delta}^2 \equiv 
\abs{\Delta^\mrm{sc}}^2 + \abs{\Delta^\mrm{pg}}^2} with
\meq{\abs{\Delta^\mrm{pg}}^2 \equiv -\sum_{\zeta,q_z;i\Omega}
\frac{g}{1+g\chi(\zeta,q_z;i\Omega)}} and
$\avg{\phi^2_{mm'}(\zeta)}_\zeta \equiv$
\meq{\frac{\sum_{\zeta,q_z,i\Omega}\phi^2_{mm'}(\zeta)
\pth{t^\mrm{sc}(\zeta,q_z;i\Omega)+t^\mrm{pg}(\zeta,q_z;i\Omega)}}
{\sum_{\zeta,q_z,i\Omega}
\pth{t^\mrm{sc}(\zeta,q_z;i\Omega)+t^\mrm{pg}(\zeta,q_z;i\Omega)}}.} 
This leads to
an expression for the self-energy 
which can be written compactly as
\meq{\Sigma_{mm'}(i\omega) = 
-G^0_N(-k_z;-i\omega)\avg{\phi^2_{mm'}(\zeta)}_\zeta
\abs{\Delta}^2.}  Together with the number equation $N =
\frac{2}{\beta}\sum_{m,i\omega}G_{mm}(i\omega)$ and
Eqs.~(\ref{LLGFunc},\ref{LLDeltaEqn}) this forms a system of equations which can
be solved self-consistently.\cite{chen_2005} 
In this way we have modified the Gor'kov theory to extend BCS theory
into the more general regime where pairing and condensation are treated
differently.

Finally, we turn to a more detailed interpretation of the ladder diagrams
and the related pair susceptibility.
In Fig.~\ref{LadderSum}, particles forming
a pair may
interact to create a new 
non-condensed pair in the same excited pair state.  That
is, we consider a pairing propagator $t_{mn}^{m'n',\mrm{pg}}
(\zeta,q_z;i\Omega)$ between states $m,n$ and $m',n'$, 
where $m = (N,p,k_z,i\omega)$ and $n = (N',p',q_z-k_z,i\Omega-i\omega)$
are pairing
partners for a given real-space configuration $\zeta$, total $z$-momentum
$q_z$, and total Matsubara frequency $i\Omega$, and
$m' = (N'',p'',k_z',i\omega')$ and $n' = (N''',p''',q_z-k_z',
i\Omega-i\omega')$ are pairing partners sharing the same $\zeta$,
$q_z$ and $i\Omega$.
Note that $m$ now includes frequency, and intra-eigenstate pairing is
not enforced here.
The bare interaction between the two pairs, as an extension of the BCS
mean-field Hamiltonian, is in a separable form $gV_{mn}^{m'n'} =
g\Delta^{0\dag}_{mn}(\zeta)\Delta^0_{m'n'}(\zeta)$.
Then the infinite particle-particle ladder summation is
\mea{&t_{mn}^{m'n',\mrm{pg}}(\zeta,q_z;i\Omega)
 =
gV_{mn}^{m'n'} - \notag \\
&gV_{mn}^{m''n''}G^0_{n''}G_{m''m'''}
t^{m'n',\mrm{pg}}_{m'''n''}(\zeta,q_z;i\Omega).}

As $V$ is separable between $m,n$ and $m',n'$, this infinite summation has
the solution
\mea{&t^{m'n',\mrm{pg}}_{mn}(\zeta,q_z;i\Omega) = \notag \\
&\frac{gV_{mn}^{m'n'}}{1 +
g\sum_{m'',m''',n''}V_{m'''n''}^{m''n''}G^0_{n''}G_{m''m'''}}.}  Using the
expression for $V$, 
and applying the intra-eigenstate pairing approximation, this
further simplifies to 
\mea{&t_{mn}^{m'n',\mrm{pg}}(\zeta,q_z;i\Omega) = \notag \\
& \frac{g\Delta^{0\dag}_{mn}(\zeta)\Delta^0_{m'n'}(\zeta)}{1+g\sum_{m'',m'''}
\phi^2_{m''m'''}(\zeta)G_{m''m'''}G^0_N(k_z-q_z;i\Omega-i\omega)}.\notag} 
This in turn
gives the $t$-matrix appearing in the self-energy as 
\meq{\sum_n t_{mn}^{m'n,\mrm{pg}}(\zeta,q_z;i\Omega) =
\phi^2_{mm'}(\zeta)t^\mrm{pg}(\zeta,q_z;i\Omega)} with
$t^\mrm{pg}(\zeta,q_z;i\Omega)$ defined in Eq.~\eqref{eq:20}. Note this also
justifies our definition of $\chi$ in Eq.~\eqref{chiEq}.

\begin{figure*}
\includegraphics[clip=true,trim = 2.25in 3.70in 1.85in 3.00in,scale=0.40]
{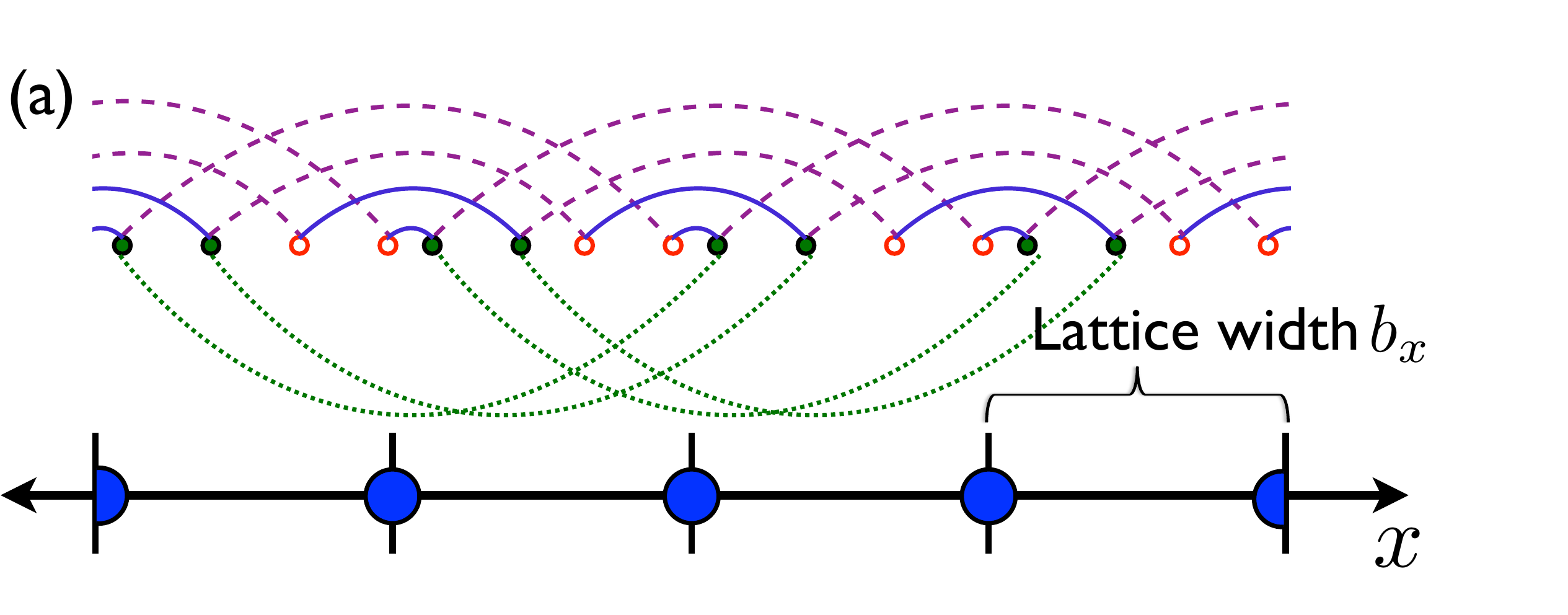}
\includegraphics[clip=true,trim = 3.5in 5.6in 4.4in 1.9in,scale=0.45]
{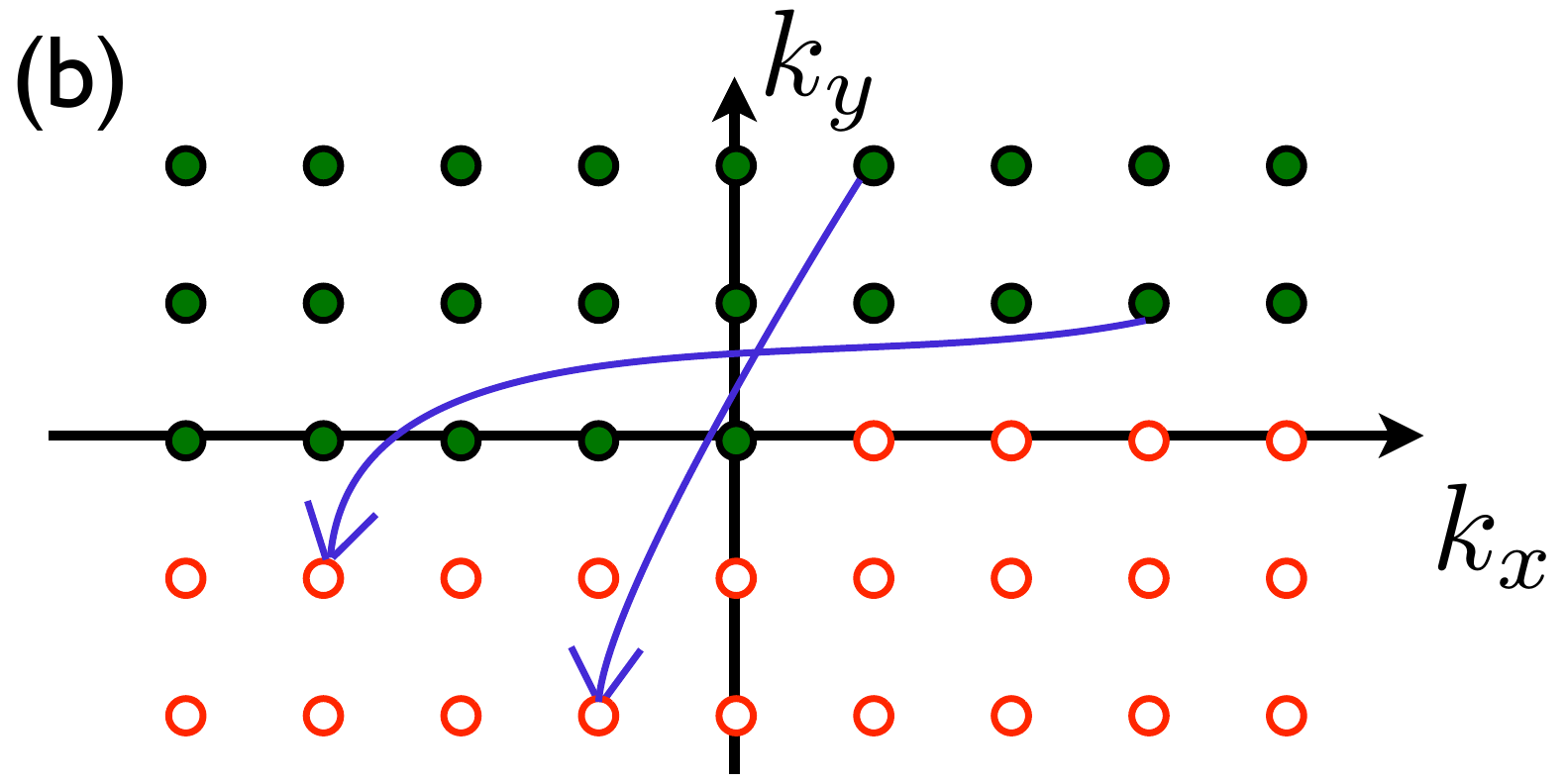}
\caption{\label{Cartoons}
(Color online) Schematic version of the two pairing models, \textbf{(a)} the
restricted ``tight-binding'' pairing for case (i), in the case of a square
lattice (see Section \ref{secNNPair}).  
The fermionic states (small open red and filled green circles) can
pair with each other only via the solid blue (closest pairing partner) and
dashed purple (next closest pairing partner) lines.  These pairs in turn connect
``nearest neighbor'' states 
via an off-diagonal Green's function matrix element (dotted
green lines; equivalent lines for the red states are omitted).  Furthermore, the
pairs can be interpreted as forming bosons at lattice sites (large blue
circles).  \textbf{(b)} The case (ii) version of pairing, in which states
(filled green and open red circles) are in a reciprocal lattice of the magnetic
translation group, and pair with states of opposite momentum index (blue
lines).
}
\end{figure*}

In summary, beyond the weak attraction limit, pairs can form
above the superconducting transition temperature.  
These pairs of electrons 
form in excited states rather than the quantum ground state.  Interpreting
the Gor'kov gap equation as a Bose Einstein condensation condition
allows us to specify a precise ladder series structure for these non-condensed pairs.
Because it will take us too far afield,
elsewhere, we discuss a precise treatment of the
parameter $\zeta$, in more microsopic detail. 
\cite{scherpelz_2012}

\section{Specifying the Pairing Basis\label{sec:pairing}}
We now turn to the specific pairing choices that can be made to solve the
system, and explore their physical effects.  We now consider only a single
$\zeta$, and for notational simplicity we now omit this parameter.
 Ensembles of $\zeta$ are discussed
elsewhere.\cite{scherpelz_2012}
In order to make further progress, we must specify 
$\Delta(\bs r)$
as well as the Landau level state 
basis we
use. This requires that one
determine the 
pairing partners of each fermionic Landau level state.  Based on
previous
work
\cite{ryan_1993,dukan_1994,vavilov_1997}
here we analyze and compare two
complementary choices for the case of an Abrikosov lattice.
Throughout we use the Landau gauge $\bs A
= (0,B\hat{\bs x},0)$ in which the energy gap of an Abrikosov lattice,
$\Delta(\bs r)$ is given by 
\mea{\Delta(\bs r) &= C\sum_m \exp\pth{i\pi
\frac{b_y}{a}m^2} \psi^\mrm{cm}_{0,mb_x,0}(\bs r) \notag \\
&= \Delta\sqrt{\frac{\sqrt{2} b_x}{L_xL_yL_z\sqrt{\pi}l_H}} \notag \\
&\ {}\times\sum_m 
\exp\bckt{i \pi
\frac{b_y}{a}m^2 + \frac{2 i m b_x y}{l_H^2}-\frac{\pth{x- m b_x}^2}{l_H^2}}}
where $\psi^\mrm{cm}_{N,X,k_z}$ is the orbit-center Landau level state for
a charge-2$e$ particle with Landau level $N$, orbit center $X$, and $z$-momentum
$k_z$. The normalization of $\Delta(\bs r)$ is chosen such that $\int d\bs r
\abs{\Delta(\bs r)}^2 = \abs{\Delta}^2$, with $L_x,\ L_y,$ and $L_z$ the sample
lengths and $l_H = \sqrt{\hbar c / eH}$ the Hall length.
Finally, the Abrikosov lattice is characterized by
unit vectors $\bs a = (0,a,0)$ and $\bs b = (b_x,b_y,0)$.  $l_H$ is related to
the unit vectors by $ab_x = \pi l_H^2$.  For a square lattice $\bs b = (a,0,0)$
while for a triangular lattice $\bs b = (\sqrt{3}a/2,a/2,0)$, and in general we
capture all Abrikosov lattices using the method in
Ref.~\onlinecite{saint-james_1969}, $\zeta =
b_y/a + ib_x/a$. Throughout the rest of this paper, we implicitly take $b_x,\
b_y,$ and $a$ to be functions of $\zeta$ through the above formula and the
restriction that $ab_x = \pi l_H^2$.

The two natural choices are to use either orbit-center wavefunctions, in which
case orbit centers positioned symmetrically about a lattice site pair together,
or to use magnetic translation group wavefunctions, in which case wavefunctions
with opposite reciprocal lattice vectors are paired.  We describe those pairs
in terms of the notation of $\Delta_{mn}^0(\zeta)$ and $\phi^2_{mm'}(\zeta)$ 
defined above.

\subsection{Orbit-center Pairing\label{sec:orb}}
One choice of pairing, originally presented by Ryan and Rajagopal is to have
fermions pair about lattice site positions in the orbit center 
basis.\cite{ryan_1993,rajagopal_1992,rajagopal_1993,rajagopal_1995}
Because the bosonic wave functions forming the Abrikosov lattice are positioned
at orbit centers $X = mb_x$, fermions which are equally spaced apart from these
positions can pair, so that the pair wavefunction
can be represented as
$\Psi^\mrm{pair}_{N_1,N_2,mb_x,Y,k_z} (\bs r) = $
\meq{\psi^\mrm{fermion}_{N_1,mb_x+Y,k_z,\upa}(\bs
r)\psi^\mrm{fermion}_{N_2,mb_x-Y,-k_z,\dna}(\bs r).\notag}  
Here
\mea{&\psi^\mrm{fermion}_{N,X,k_z}(\bs r) = \sqrt{\frac{1}{L_y L_z 2^N
N!}}\pth{\frac{1}{\pi l_H^2}}^{1/4} \notag \\
&\ {}\times\exp\pth{i k_z z+\frac{i X y}{l_H^2}-
\frac{\pth{x-X}^2}{2l_H^2}}H_N\pth{\frac{x - X}{l_H}}} where $H_N$ is the
$N$\textit{th}-order Hermite polynomial.  The $\Delta_{mn}^0(\zeta)$ 
and associated
factors are calculated in Appendix \ref{RRAppendix}. 
Quite generally (presuming inter-eigenstate
pairs), the result is
\begin{widetext}
\meq{\Delta^0_{m = (N_1,X+Y,k_z),n=(N_2,X-Y,-k_z)}(\zeta) =\begin{cases} \sqrt{\frac{
b_x}{L_xL_y^2L_z^2\pi l_H^2}} \frac{1}{2^{N_1+N_2}\sqrt{N_1!N_2!}}  e^{i\pi
(b_y/a)(X/b_x)^2}e^{-Y^2/l_H^2} & \mrm{if } X = mb_x, m \in \mbb{Z}, \\ 0 &
\mrm{otherwise.} \end{cases}}
\end{widetext}

\subsection{Magnetic Translation Group Pairing}
The other choice of pairing, originally presented by Dukan, Andreev, and
Tesanovic\cite{dukan_1991,dukan_1994} 
and in parallel work by Akera, MacDonald, Norman, and
Girvin,\cite{akera_1991,norman_1992,norman_1995} and Nicopoulos and 
Kumar,\cite{nicopoulos_1991} is to use the 
magnetic translation group (MTG) for the
fermions, with an index $\bs k$.\footnote{Note that
Refs.~\onlinecite{dukan_1991,dukan_1994} 
use a different gauge than we use here.}
Following Ref.~\onlinecite{dukan_1994} we 
choose a unit lattice site for the magnetic
translation group, which must be twice the size of the Abrikosov lattice site,
with unit vectors $2\bs a$ and $\bs b$.  Then the reciprocal unit vectors are
$\bs a^* = (-b_y/l_H^2,b_x/l_H^2)$ and $\bs b^* = (2a/l_H^2,0)$ so that $\bs a_i
\bs a_j^* = 2\pi\delta_{ij}$ where $\bs a_i$ are the unit vectors and $\bs
a_j^*$ are the reciprocal unit vectors.  Restricting $\bs k = (k_x,k_y)$ 
to be within the
limits of the cell $(\bs a^*, \bs b^*)$ gives a complete set of functions.
\cite{bychkov_1983}  We also note that $\psi$ is now dependent on $\zeta$, the
configuration of the Abrikosov lattice.

This pairing occurs between opposite $\bs k$.  To conform with
our assumption of diagonal pairing, we
need to specify that $N_1 = N_2$, and then \meq{\Psi^\mrm{pair}_{N,\bs
k}(\bs r) = \psi^\mrm{fermion}_{N,\bs k,k_z,\upa}(\bs
r)\psi^\mrm{fermion}_{N,-\bs k,-k_z,\dna}(\bs r)\notag} with
\begin{widetext}
\mea{\psi^\mrm{fermion}_{N,\bs k,k_z} &=  \sqrt{\frac{1}{2^N
N!}}\pth{\frac{1}{\pi l_H^2}}^{1/4}\sqrt{\frac{b_x}{L_xL_yL_z}} \exp\pth{i k_z
z} \sum_m \exp\pth{i\frac{\pi b_y}{2a}m^2 + i m k_x b_x} \nonumber \\ 
&\ {}\times
\exp\brc{i\pth{k_y+\frac{\pi m}{a}}y - \frac{\bckt{x - \pth{k_y + \frac{\pi
m}{a}}l_H^2}}{2 l_H^2}}H_n\brc{\bckt{x - \pth{k_y + \frac{\pi
m}{a}}l_H^2}/l_H}.} 
\end{widetext}
The $\Delta^0_{mn}(\zeta)$ and
associated factors are calculated in 
Refs.~\onlinecite{dukan_1994,vavilov_1997}, 
with the result that for the
lowest Landau level
\mea{\Delta&{}^0_{m = (0,\bs k,k_z),n = (0,-\bs k,-k_z)}(\zeta) = 
\sqrt{\frac{b_x}{L_xL_yL_z\sqrt{2\pi}l_H}}
 \notag \\
&\ {}\times \exp\pth{ -\pth{k_yl_H}^2}
\theta_3\left(\bckt{-k_x+ik_y}b_x\bigg|\frac{-b_y}{a}+\frac{i\pi
l_H^2}{a^2}\right),}  
where $\theta_3(u | \tau) =
\sum_{n = -\infty}^\infty \exp\pth{2inu + i \pi n^2 \tau}$ is the third elliptic
theta or Jacobi theta function.
Further $\Delta_{mn}^0(\zeta)$ for higher Landau levels
can be iteratively calculated from this.\cite{dukan_1994}

\section{Nearest Neighbor Pairing Approximation for Real Space
Pairing\label{secNNPair}}
While the MTG method for pairing results in each fermion pairing with exactly
one other degenerate eigenstate, that is not the case for
orbit center pairing.  If $X$ is a lattice site, then a fermion at $X + Y$ can
pair not only with $X - Y$, but also with $X - Y + 2nb_x$ with $n \in \mbb{Z}$.
A very complicated matrix thus results for the Green's function, but here we
demonstrate that this matrix can be substantially simplified such that it is
analytically tractable yet incorporates the important physics.

We begin by noting that $\Delta^0_{mn}$ is proportional to $\exp(-Y^2/l_H^2)$
where $Y$ is half of the distance between the two states that pair.  Thus,
pairing between states that are well separated will be exponentially suppressed.
Using only a single pairing partner is inadequate, however, because it fails to capture
interference effects between two pairing partners, as will be demonstrated
below.  Using two pairing partners does capture these effects and
reproduces the correct local density of states for the lattice.  This is
reminiscent of a hierarchical
``tight-binding'' scheme, in that the nearest pairing partner
can be considered an ``on-site'' interaction, and the second-nearest pairing
partner allows for interactions between lattice sites.

To implement this approach, we begin with Eq.~\ref{LLGFunc}, $G_{mm'}(i\omega) = $
\meq{G^0_{m}(i\omega)\delta_{mm'} -
G^0_m(i\omega)\sum_{ln}\Delta_{ml}G^0_{l}(-i\omega)\Delta^\dag_{ln}
G_{nm'}(i\omega).
\tag{\ref{LLGFunc}}}
This can be further simplified by separately including the Green's functions of
pairing partners, resulting in a new equation for the Green's functions
$G_{mm'}(i\omega) = $
\meq{G^0_{m}(i\omega)\delta_{mn} + 
G^0_m(i\omega)\sum_n\Delta_{mn}[-\tilde G_{nm'}(i\omega)].\label{GNNEq}}
  Here
by $\tilde G$ we mean (a) flipping the sign of $i\omega$ and (b) conjugating
the $\Delta$ appearing within $G$. By this definition, we also have that
$\tilde G_{mm'}(i\omega) = $
\meq{G^0_m(-i\omega)\delta_{mm'} + G^0_m(-i\omega)
\sum_{n}\Delta^\dag_{mn}G_{nm'}(i\omega).}
This correctly reproduces the original Green's function above, and 
by inverting Eq.~\eqref{GNNEq}, we have that 
\begin{widetext}
\meq{G^{-1}(i\omega) = 
\pmt{\ddots & \ddots & & & & \\
\ddots & \ddots & \Delta^\dag_{X-Y,X+Y} & 0 & 0 & 0 \\
& \Delta_{X-Y,X+Y} & [G^0_{X+Y}(i\omega)]\inv 
& \Delta_{X+2b_x-Y,X+Y} & 0 & 0 \\
& 0 & \Delta^\dag_{X+2b_X-Y,X+Y} & -[G^0_{X+2b_x-Y}(-i\omega)]\inv & 
\Delta^\dag_{X+2b_x-Y,X+2b_x+Y} &
0 \\
& 0 & 0 & \Delta_{X+2b_x-Y,X+2b_x+Y} & [G^0_{X+2b_x+Y}(i\omega)]\inv & 
\Delta_{X+4b_x-Y,X+2b_x+Y} & \\
& 0& 0& 0& \Delta^\dag_{X+4b_x-Y,X+2b_x+Y} & \ddots  & \ddots \\
& & & & & \ddots  & \ddots}.}
\end{widetext}
Here we use only the index corresponding to the orbit center, which is the only
one which varies throughout this matrix. We also take $Y > 0$;  if $Y < 0$ 
positions in the matrix will flip but it will be otherwise unchanged. Also note
that the values of $G_0^{-1}$ are actually independent of orbit center; the
indices remain for clarity.

This Green's function matrix is computationally simplified as compared to the full
orbit-center pairing scheme. Furthermore, in the case of a
square or triangular lattice, its eigenvalues can be found analytically.  For
both a square lattice ($\zeta = i$) and a triangular lattice ($\zeta = 1/2 +
\sqrt{3}i/2$), we have the property that $\Delta^0_{X-Y+2b_x,X+Y+2b_x} =
\Delta^0_{X-Y,X+Y}$, implying that the entries in $G^{-1}$ repeat with
period two along the diagonal.  
We can then posit a plane-wave solution for the eigenvectors, with $a$ and $b$
constants: 
\meq{\bs v = \sum_q e^{i q k_j}\pth{a\kt{X_{\bar{m}}+2qb_x} + b\kt{X_m+2qb_x}},}
where for compactness we set $\bar m = X-Y$, $m = X+Y$, and $\bar n =
X-Y+2b_x$, using the symmetry properties of $\Delta^0_{mn}$.
We obtain 
\meq{G^{-1}\bs v = \pmt{ \vdots \\ \Delta_{\bar{m}m}b + (\omega-\xi)a +
\Delta_{\bar n m} e^{i k_j} b \\
\Delta^\dag_{\bar n m} a + (\omega + \xi) e^{i k_j} b + \Delta^\dag_{\bar m m}
e^{i k_j} a \\ \vdots}.}

The energy eigenvalues of this system will be zero eigenvalues of
$G^{-1}$.  For a nontrivial solution, we obtain
$
-\Delta^\dag_{\bar n m}\pth{\Delta_{\bar m m}e^{-i k_j}  + \Delta_{\bar n m}} 
+ (-\omega^2-\xi^2) 
- \Delta^\dag_{\bar m m} \pth{\Delta_{\bar m m} + \Delta_{\bar n m}e^{i k_j}} 
 = 0.$ 
We thus have $-\omega^2 =$
\meq{ \xi^2 + \abs{\Delta_{\bar mm}}^2 + \abs{\Delta_{\bar nm}}^2 +
e^{-i k_j} \Delta_{\bar m m}\Delta^\dag_{\bar nm} + e^{i k_j} \Delta^\dag_{\bar
mm}\Delta_{\bar nm}.}

Here $\Delta_{\bar mm} = C_{\bar mm} 
e^{i \pi (b_y / a)(X/b_x)^2}$, where $C_{\bar mm}$ is a real
number, and $\Delta_{\bar nm} = C_{\bar nm} 
e^{i \pi (b_y / a)[(X+1)/b_x]^2}$.  We further
simplify 
\mea{&
e^{-i k_j} \Delta_{\bar m m}\Delta^\dag_{\bar nm} + e^{i k_j} \Delta^\dag_{\bar
mm}\Delta_{\bar nm} \notag \\ 
&= 2\trm{Re}\pth{e^{-i k_j} \Delta_{\bar m
m}\Delta^\dag_{\bar nm}} \notag \\
&= 2C_{\bar mm}C_{\bar nm}\trm{Re}\pth{e^{-i
k_j}e^{\pm i\pi(b_y/a)}}\notag} where we have used the $b_x$ and $b_y$ specific to a
square or triangular lattice, and the $+$ is for $X/b_x$ an odd integer, $-$ for
$X/b_x$ an even integer.
We then have $-\omega^2 = $
\meq{\xi^2 + \abs{\Delta_{\bar mm}}^2 + \abs{\Delta_{\bar nm}}^2 +
2C_{\bar mm}C_{\bar nm} \trm{Re}\pth{e^{-i k_j} e^{\pm i \pi (b_y/a)}}.}  We
also note that for a system of $L_x = 2N_xb_x$, we must have $N_x k_j = 2\pi j$
to enforce periodic boundary conditions, giving $k_j = 2\pi j / N_x$ with $j$
ranging from $0$ to $N_x-1$.  With this approximation in place, we may turn to
the results of these pairing theories.

\section{Results and Discussion\label{secResults}}
\begin{figure}
\includegraphics[clip=true,trim = 2.30in 3.85in 2.45in 3.95in,scale=0.9]
{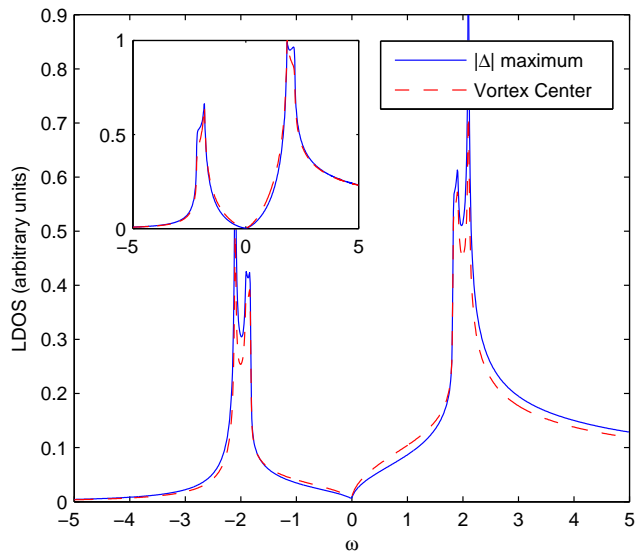}
\caption{\label{LDOSomega}
(Color online) 
Plots of the local density of states $N(\bs r;\omega)$ vs. $\omega$ on a square
lattice, for $\Delta^\mrm{sc} = 2$ 
and $\delta = 0.01$, and normalized to a maximum LDOS
of $1.0$.  \textbf{(Main figure)} Case (i), using the
nearest neighbor approach for calculations with $L_x = 4b_x$. \textbf{(Inset)} 
Case (ii).
}
\end{figure}

\begin{figure}
\includegraphics[clip=true,trim = 2.15in 4.43in 2.15in 4.43in,scale=0.82]
{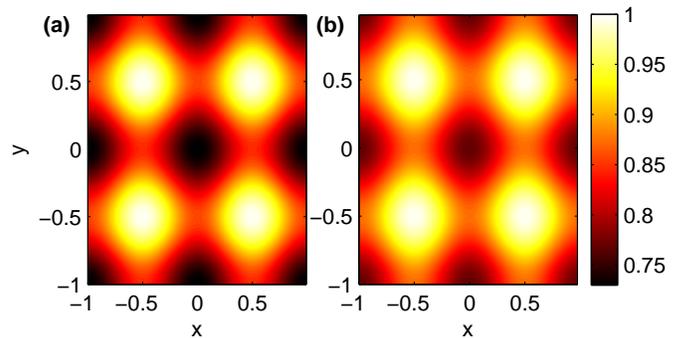}
\caption{\label{LDOSTopo}
(Color online) 
Plots of the local density of states $N(\bs r;\omega)$ for \textbf{(a)} case (i)
and \textbf{(b)} case (ii), normalized to the local density of states at the
vortex cores (located at $(x,y) = (\pm 0.5,\pm 0.5)$) and as a function of
position on a square lattice, for $\omega = 0.4$, $\Delta^\mrm{sc} = 2$, 
and $\delta =
0.05$ (slightly higher than Fig.~\ref{LDOSomega} to permit faster computation
without a substantial change in accuracy).
}
\end{figure}

This paper has been rather extensively devoted to theoretical
formalism which characterizes the fermionic degrees of freedom.
Among the most direct experimentally relevant consequences is
the behavior of the local density of states (LDOS)
$N(\bs r;\omega)$ vs. $\omega$. 
Here we address
this density of states in the very low temperature regime.
This is
experimentally accessible using scanning tunneling 
microscopy.\cite{fischer_2007}  The local density of states $N(\bs
r;\omega)$  is calculated via
$N(\bs r;\omega) = 2\trm{ Im }G^\mrm{ret}(\bs r,\bs r;\omega)$.
We determine
$G^\mrm{ret}(\bs r,\bs r';\omega) =
\sum_{mm'}G_{mm'}(\omega+i\delta)\psi_m(\bs r)\psi^\dag_{m'}(\bs r')$ in the
limit $\delta \ria 0^+$, and for simplicity we address the square lattice 
at the lowest temperature where there is
only a condensate.  
Also for simplicity our illustrative calculations are for the limiting case of
the lowest Landau level, an $s$-wave gap, $N_x = 2$ for orbit center pairing, 
and normalized to set the mass
$m = 1$.

It has been argued quite generally that
$N(\bs r;\omega)$
exhibits a gapless behavior,
in contrast to superconductivity without a magnetic
field. 
This
gaplessness is due\cite{maniv_2001} to the fact that all
fermions are delocalized,
unlike in the vortex cores of the low field limit.
This observation has direct application to magnetic oscillation
measurements, as a gapped state would dampen these oscillations significantly.
\cite{maniv_2001}
More specifically in the MTG pairing scheme, 
systematic studies in
Ref.~\onlinecite{dukan_1994} show that the energy
eigenvalues satisfy  
$E_{N,\bs k,k_z} = \xi_{N,k_z}^2+\abs{\Delta_{m = (N,\bs k,k_z),n = (N,-\bs
k,-k_z)}}^2$.  Because $\Delta_{mn}$ always features a zero for the Abrikosov
lattice, $E$ will as well.

The situation for orbit center pairing is more complex, but with the nearest
neighbor pairing scheme above, we are able to analytically demonstrate
gaplessness in this case as well as for the square and triangular Abrikosov
lattices.  Such gaplessness will occur when 
$Y = 0.5b_x$, which means the magnitudes of
$\abs{\Delta_{\bar mm}}^2 + \abs{\Delta_{\bar nm}}^2$ and $2\abs{\Delta_{\bar
nm}\Delta^\dag_{\bar mm}}$ are equal (using the notation from the previous
section).  Thus, for gaplessness to occur, we must
also have that 
\meq{\trm{arg }\pth{e^{-i k_j}e^{\pm i \pi (b_y/a)}} =
\pi,} or
$j_\mrm{gapless} = \frac{N_x}{2}\pth{1\pm\frac{b_y}{a}.}$  For a square lattice,
any  $N_x = 2m,\ m \in \mbb{N}$ will thus feature a gapless state, while for 
a triangular lattice, any $N_x = 4m$ will feature a gapless state.  This also
shows the importance of including ``nearest-neighbor'' pairs; neglecting them
cannot produce gapless states.

Beyond this analytical assessment, we observe additional similarities and
differences between the pairing schemes, and gain intuition about 
experiment.
Figure \ref{LDOSomega} presents a plot of the local density of states as a function
of energy for the
two different cases and for two different positions of the probe: one
at the vortex center (dashed) and the other at the point of maximum $|\Delta|$
(solid curve). 
Again prominent in the features of these LDOS plots is that both cases show
gapless behavior.  
In both cases, there is only one ``nodal''
state $m$ per lattice site at which its total excitation $E = 0$.  This produces
the distinctive parabola-like shape in Fig.\ \ref{LDOSomega}, which still
touches $N(\bs r;0) = 0$ but does not exhibit a complete gap away from $\omega =
0$. 
The
variation between the solid and dashed curves is rather small, also reflecting
this point.

Figure \ref{LDOSTopo} presents a contour plot of
$N(\bs r;\omega)$ as a function of $\bs
r$.  This figure nicely illustrates that
the real space and reciprocal lattice space pairing schemes
are rather similar here. Both reflect the symmetry of the Abrikosov lattice.
However the former shows slightly more contrast than the latter.
\footnote{Topographic plots at higher frequencies (not
shown) have slight deviations from an Abrikosov lattice symmetry for case (i), 
due to
the small $N_x = 2$. Similarly,
the ``kinks'' in the low-frequency LDOS are likely due to the
same approximation.}

\section{Conclusions\label{secConclusion}}

It is anticipated that the formalism in this paper will be relevant
to both ultracold gases in the BCS-BEC
crossover regime, 
and possibly to the pseudogap phase in high temperature superconductors.
Magnetic field effects in the latter have revealed a number
of mysteries, which appear to be associated with a normal state
pairing gap.
For the cold Fermi gases there is considerable interest
in effects
arising from artificial gauge fields or rapid rotation.
Previous work on these Fermi gases
\cite{Horotation,*Radzihovskyrotation,*Cooperrotation}
has presumed, rather improbably,
that even in the BEC regime, pairing and condensation
appear at the same temperature.
Also important is a better understanding of the normal state
and of how condensation can take place
if the superconductor or superfluid is effectively one dimensional.

This paper has focused on the nonlinear gap
region in the presence of a magnetic field.  A next step is to address
calculations of the onset of superfluid coherence at temperature $T_c(H)$, which
is taken to be less than the onset of pairing, $T^*(H)$, in contrast to
previous work.\cite{Horotation,*Radzihovskyrotation,*Cooperrotation,
rasolt_1992,ryan_1993,rajagopal_1991,rajagopal_1992,dukan_1991,dukan_1994}
To this end, our work has established that the Gor'kov equations lend themselves
to the nonlinear, analytic approach required, provided only degenerate energy
states are paired.  It has also shown that unique pairing partners are not
required for a tractable theory.  A robust result of this theory is that gapless
states are present for both real and momentum-space
pairing theories in a very high field.  

In summary,
it is hoped that this formalism lays the foundation to explore a variety of
magnetic field
effects in the pseudogap phase and throughout the
BCS-BEC crossover.

We thank Victor Gurarie, Tin-Lun Ho, and Vivek Mishra for helpful
discussions.  This work is supported by NSF-MRSEC Grant 0820054. P.S.
acknowledges support from the Hertz Foundation.

\appendix

\section{Details of the Orbit-Center Pairing Calculations \label{RRAppendix}}
Since $\Psi^\mrm{pair}$ is independent
of $k_z$, that subscript will be dropped in pair terms that follow.

Ref.~\onlinecite{ryan_1993} provides an identity, 
\mea{&\psi_{N_1,X+Y}(\bs r_1)\psi_{N_2,X-Y}(\bs r_2) = \sum_{P =
0}^{N_1+N_2}C^P_{N_1N_2} \notag \\
&\ {}\times \psi^\mrm{cm}_{P,X}(\bs r_\mrm{cm})
\psi^\mrm{rel}_{N_1+N_2-P,2Y}(\bs r_\mrm{rel})} 
where $N_1,N_2,$ and $P$ are
Landau levels, $\bs r_{cm} = (\bs r_1 + \bs r_2)/2$, $\bs r_\mrm{rel} = \bs r_1
- \bs r_2$, $\psi^\mrm{cm}$ is an orbital function with $l \ria 2^{-1/2} l$
(appropriate for a charge-2e particle), $\psi^\mrm{rel}$ has $l \ria 2^{1/2} l$,
and $C^P_{N_1N_2}$ is a complicated combinatorial prefactor which is equal to 1
if $N_1 = N_2 = 0$.  Thus, we know that 
\mea{&\Psi^\mrm{pair}_{N_1,N_2,X,Y} (\bs
r) = \sum_{P = 0}^{N_1+N_2} C^P_{N_1N_2}\psi^\mrm{cm}_{P,X}(\bs
r)\psi^\mrm{rel}_{N_1+N_2-P,2Y}(0) \notag \\ 
&= \sum_{P = 0}^{N_1+N_2}
C^P_{N_1N_2}\sqrt{\frac{1}{L_y L_z 2^{N_1+N_2-P} (N_1+N_2-P)!}}\notag \\
&\ {}\times\pth{\frac{1}{2
\pi l_H^2}}^{1/4} e^{-Y^2/l_H^2}\psi^\mrm{cm}_{P,X}(\bs r) \notag }

We can now proceed to calculate the $\Delta_{mn}$ elements for all possible
pairs, with the result that 
\begin{widetext}
\mea{
&\Delta_{m=(N_1,X+Y),n=(N_2,X-Y)} = \int d\bs r \Delta(\bs r)
\psi^\mrm{fermion\dag}_{N_1,X+Y,k_z}(\bs r)
\psi^\mrm{fermion\dag}_{N_2,X-Y,-k_z}(\bs r) \notag \\ 
&=
\Delta\sqrt{\frac{\sqrt{2} b_x}{L_xL_yL_z\sqrt{\pi}l_H}} \int d\bs r
\sum_{m}\exp\pth{i\pi \frac{b_y}{a}m^2} \psi^\mrm{cm}_{0,mb_x}(\bs r) 
\notag \\
&\ {}\times
\sum_{P = 0}^{N_1+N_2}C^P_{N_1N_2}\sqrt{\frac{1}{L_yL_z 2^{N_1+N_2-P}
(N_1+N_2-P)!}} \notag 
\pth{\frac{1}{2 \pi l_H^2}}^{1/4}
e^{-Y^2/l_H^2}\psi^\mrm{cm\dag}_{P,X}(\bs r) \\ 
&= \begin{cases}
\Delta\sqrt{\frac{\sqrt{2} b_x}{L_xL_yL_z\sqrt{\pi}l_H}} e^{i\pi
(b_y/a)(X/b_x)^2}
C^0_{N_1N_2}\sqrt{\frac{1}{L_y L_z 2^{N_1+N_2}
(N_1+N_2)!}}\pth{\frac{1}{2 \pi l_H^2}}^{1/4} e^{-Y^2/l_H^2} & \mrm{if } X =
mb_x, m \in \mbb{Z}, \\ 0 & \mrm{otherwise.} \end{cases}\notag }
Here $C^0_{N_1N_2} =
\sqrt{\frac{(N_1+N_2)!}{N_1!N_2!}}\frac{(-1)^{N_2}}{2^{(N_1+N_2)/2}}$ to give
\meq{\Delta_{m=(N_1,X+Y),n=(N_2,X-Y)} =\begin{cases} \Delta\sqrt{\frac{
b_x}{L_xL_y^2L_z^2\pi l_H^2}} \frac{1}{2^{N_1+N_2}\sqrt{N_1!N_2!}}  e^{i\pi
(b_y/a)(X/b_x)^2}e^{-Y^2/l_H^2} & \mrm{if } X = mb_x, m \in \mbb{Z}, \\ 0 &
\mrm{otherwise.} \end{cases}}
\end{widetext}

\bibliography{Review3,Review4}

\end{document}